# Optical-Cavity-Induced Current

Garret Moddel,* Ayendra Weerakkody, David Doroski and Dylan Bartusiak
Department of Electrical, Computer, and Energy Engineering, University of Colorado, Boulder, Colorado, 80309-0425, USA

The formation of a submicron optical cavity on one side of a metal–insulator–metal (MIM) tunneling device induces a measurable electrical current between the two metal layers with no applied voltage. Reducing the cavity thickness increases the measured current. Eight types of tests were carried out to determine whether the output could be due to experimental artifacts. All gave negative results, supporting the conclusion that the observed electrical output is genuinely produced by the device. We interpret the results as being due to the suppression of vacuum optical modes by the optical cavity on one side of the MIM device, which upsets a balance in the injection of electrons excited by zero-point fluctuations. This interpretation is in accord with observed changes in the electrical output as other device parameters are varied. A feature of the MIM devices is their femtosecond-fast transport and scattering times for hot charge carriers. The fast capture in these devices is consistent with a model in which an energy $\Delta E$ may be accessed from zero-point fluctuations for a time $\Delta t$, following a $\Delta E \Delta t$ uncertainty-principle-like relation governing the process.

*Corresponding author: moddel@colorado.edu

## 1. Introduction

Metal–insulator–metal (MIM) tunnel diodes have been used to provide rectification and nonlinearity [1–3] for a variety of applications. The insulator forms a barrier that charge carriers—electrons or holes—must cross to provide current when a voltage is applied across the device. In addition, current can be produced by the direct absorption of light on one of the metal surfaces of an MIM sandwich structure, which generates the hot carriers that cross the metal and are injected into the insulator. This internal photoemission [4,5] is also called photoinjection. For current to be provided, the metal layer must be thinner than the hot-carrier mean-free path length so that the carriers can cross it without being scattered. Once they reach the insulator, the hot carriers must have sufficient energy to surmount the energy barrier at the interface and traverse the insulator ballistically above its conduction band edge, or alternatively they can tunnel through the insulator. Thinner insulators favor tunneling [4,6]. After entering the metal base electrode on the other side, the hot carriers are scattered and captured.

For over two decades, our lab has designed and fabricated MIM diodes for ultrahigh speed rectification [3]. We have found that incorporating a thin optical resonator used as an optical cavity on one side of the MIM structure induces a reduction in the device conductivity measured over a range of several tenths of a volt [7]. At lower, submillivolt voltages, a persistent induced current and voltage is evident, which we report here. We describe observed trends in the current output as the thicknesses of the optical and electrical layers in the device are varied. After showing these trends, we present the results of a range of tests that are carried out to check whether the results could be due to some sort of an experimental artifact. Finally, to explain what could produce the observed output, we present a conjectural photoinjection model in which hot carriers are excited by the quantum vacuum field.

The devices consist of thin optical cavities deposited over MIM structures, as depicted in Figure 1. The cavity thickness is in the range of tens of nanometers up to approximately 1 μm, which results in a cavity optical mode density with a wavelength dependence described by an Airy function [8]. This cavity largely suppresses wavelengths longer than twice the cavity thickness multiplied by the refractive index of the transparent dielectric. For our devices, the resulting wavelength cutoff, above which modes are suppressed, varies from the near-infrared (NIR) through the near-ultraviolet, depending on the cavity thickness. The MIM structures include a nanometer-thick insulator to form the barrier. The upper electrode is sufficiently thin to allow hot carriers that are photoexcited on the optical cavity side of the electrode to penetrate the electrode and reach the insulator without being scattered.

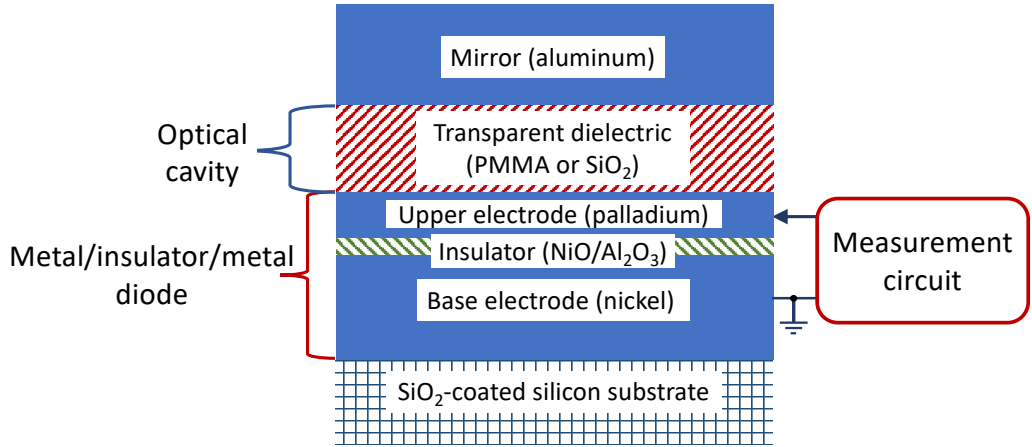

**Figure 1.** Device cross section, showing a metal–insulator–metal (MIM) structure adjoining an optical cavity. The electrical characteristics of the device are measured between the two metal layers of the MIM structure, where the polarity of the upper electrode voltage is with reference to the base electrode, which is defined as ground. Positive current is defined to be in the direction of the arrow.

## 2. Materials and Methods
*2.1 Device Fabrication*

Two different processes were used to form the devices. Submicron devices were fabricated using a germanium shadow-mask (GSM) process [9,10]. Using a deep-ultraviolet stepper, a 250 nm wide germanium bridge is formed over an $SiO_2$-coated surface of a silicon wafer, as depicted in Figure 2a. First the nickel base electrode is evaporated under the bridge from one side. This is followed by native $NiO_x$ growth at room temperature, and then by conformal $Al_2O_3$ deposited by sputtering. After the insulator is formed, the palladium upper electrode is evaporated from the opposite side. The resulting overlap of the two metals forms an ellipse with an area of 0.02 ± 0.006 μm², as shown in Figure 2b. After the germanium bridge is removed, a transparent dielectric, i.e., spun-on polymethyl methacrylate (PMMA) or sputtered $SiO_2$, is deposited to form an optical cavity over the MIM structure. The dielectric layer is then coated with an aluminum mirror. In addition to providing a reduced density of optical modes, the optical cavity encapsulates and stabilizes the MIM structure, blocking further oxidation. It should be noted that the use of PMMA to support the germanium bridge during fabrication, as shown in Figure 2a, is independent of whether PMMA is also used in a subsequent layer to form an optical cavity.

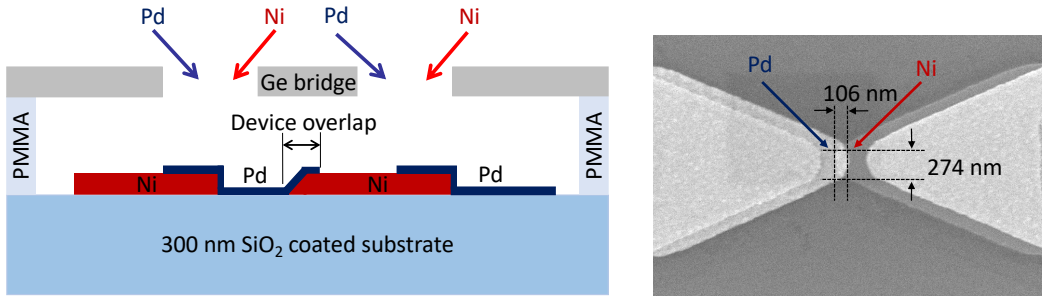

**Figure 2.** Germanium shadow mask (GSM) device fabrication. (**a**) Depiction of fabrication process, showing a cross-sectional view of materials deposited under a germanium bridge. The $NiO_x$ and $Al_2O_3$ insulating layer formed over the Ni layer is not shown. The active area of the device is formed in the overlap region. (**b**) A scanning electron microscope (SEM) image of a completed device, with an overlap area of 0.02 ± 0.006 μm². The Ni and Pd-coated regions are indicated; the lightest regions, in the center and at the left and right-hand sides, are coated with both Ni and Pd layers with the insulator layer between them.

Additional devices, having larger areas, were fabricated using standard photolithographic techniques. The top view of one of these devices is shown in Figure 3.

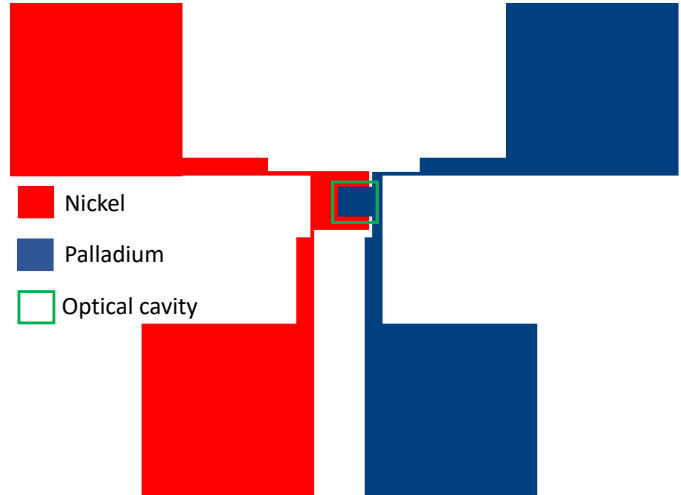

**Figure 3.** Photolithographic device. Top view of devices formed by the standard photolithographic technique. The overlap of the pallidum upper electrode, shown to the right, with the nickel lower electrode, shown to the left, forms active square regions with edge lengths between 5 and 100 μm.

The thicknesses and refractive indices of most of the dielectric layers were determined by UV–Vis–NIR variable angle spectroscopic ellipsometry (VASE) measurements. The measured refractive indices for the spun-on PMMA and deposited $SiO_2$ are 1.52 and 1.49, respectively, at a wavelength of 300 nm. The $Al_2O_3$ thickness values were measured by VASE on silicon witness samples that were placed in the sputtering system along with the devices. For the photolithographic devices, we also measured the native $NiO_x$ thickness by VASE and found it to be 2.3 nm. In addition, a monolayer (0.4 nm) of photoresist remained over the $NiO_x$ layer in the photolithographic devices. The reason for this is that after depositing the insulator and patterning the upper electrode using a liftoff process, we could not use the standard oxygen plasma to clean residual photoresist off the insulator surface without further oxidizing the insulator, with the result that the devices become too resistive. For GSM devices the patterning was accomplished by the shadow-mask deposition depicted in Figure 2a, and so no photoresist was required. Because the thickness of the native $NiO_x$ insulator in the GSM MIM structures could not be measured directly, we determined its effective thickness from electrical measurements and simulations of Ni/native $NiO_x$/Pd structures. We extracted a $NiO_x$ thicknesses of 0.6–1 nm for effective barrier heights in the range of 0.06–0.08 eV [11]. The barrier height was calculated by performing a Fowler–Nordheim analysis on a low resistance (~100 Ω) device with nonlinear current–voltage characteristics. The native NiOx thickness for GSM structures is smaller than that for the photolithographic devices because of the higher processing temperatures for the photolithographic devices, and also possibly because the junctions in GSM structures were partially protected by the germanium bridge. The total effective insulator thickness for the $Al_2O_3$/$NiO_x$ combination is the sum of the thickness values for each layer. Thickness values for the Ni base electrodes are 38 nm for the GSM devices and 50 nm for the photolithographic devices. The aluminum mirror is 150 nm thick for all devices. The thickness values for the other layers in the devices for each figure are provided in Table 1.
.

Table 1. Device parameters for each figure.

| Figure | Area (μm²) | NiO$_x$ (nm) | Al$_2$O$_3$ (nm) | Pd upper electrode (nm) | Transparent dielectric (nm) | material |
|---|---|---|---|---|---|---|
| 4(a) | 0.02 | | 1.3 | 8.3 | 33-1100 | PMMA |
| 4(b) | 0.02 | 1 | 0.9 | 8.3 | 33-1100 | PMMA or SiO$_2$ |
| 5(a)[1] | 10,000 | 2.3 + 0.4 resist | 2.3 | 8.7-24 | 11 | SiO$_2$ |
| 5(b)[1] | 625 | 2.3 + 0.4 resist | 0.7-1.5 | 15 | 11 | SiO$_2$ |
| 6(a) | 0.02 | 1 | 0.7 | 8.3 | 35 | PMMA |
| 6(b) | 25-10,000 | 2.3 + 0.4 resist | 2.3 | 12 | 11 | SiO$_2$ |
| 7(b) | 0.02 × 16 | 1 | 0.9 | 15.6 | 107 | PMMA |
| 8 | 0.02 | 1 | 0.9 | 8.3 | 36 | PMMA |
| 8 | 0.02 | 1 | 0.7 | 8.3 | 50 | SiO$_2$ |
| 9 | 0.02 | 1 | 0.7 | 8.7 | - | PMMA |
| 10 | 0.02 | 1 | 05 | 8.7 | 33 | PMMA |
| 11 | 0.02 | 1 | 0.7 | 8.7 | 35 | PMMA |

[1] Photolithographic devices; all other devices used the GSM fabrication process.

## 2.2 Device Measurement

Once the MIM structures were fabricated, we carried out current–voltage ($I(V)$) measurements at room temperature using a four-point probe configuration to circumvent the effects of lead resistance. A high precision Keithley 2612 source meter (calibrated to NIST standards) was used to source either voltage or current across two pads, and an HP 3478A digital multimeter (DMM) was used to measure the voltage drop across the MIM junctions. Although the standard technique is to source a voltage and measure the current using the source meter, this can result in erroneous offsets for low resistance devices, e.g., for currents on the order of 1 nA through a device having a resistance less than 1 MΩ. Therefore, we carried out some of the measurements, particularly for low-resistance devices, by sourcing the current (±0.06% ± 100 pA accuracy) and measuring the voltage. To eliminate any effects due to thermoelectric potentials resulting from a temperature difference between the source meter and the probes, we used a current reversal method [12]. Following this method, we performed two measurements with currents of opposite polarity, i.e., one when the base electrode was grounded and another when the upper electrode was grounded, and then we subtracted the difference in the currents to yield the final value.

In fabricating and testing tens of thousands of MIM devices, we generally found a wide range of resistances for nominally the same fabrication conditions due to slight uncontrollable variations in the insulator thickness [13], which is fewer than 10 lattice constants thick. In most cases, the measurement results presented are averages across each wafer chip, with error bars showing the standard deviation.

## 3. Results
### 3.1 Electrical Response Measurements
The electrical response of MIM devices having an adjoining optical cavity is shown in Figure 4. In Figure 4a, the $I(V)$ curves extend into the second quadrant and therefore exhibit a positive power output. For linear $I(V)$ characteristics, the maximum power is $|I_{SC}V_{OC}|/4$, where $I_{SC}$ and $V_{OC}$ are the short-circuit current and open-circuit voltage, respectively. For the device with a 33 nm thick cavity, the maximum power is 1.4 pW in a 0.02 μm$^2$ area (see the SEM image of Figure 2b). The short-circuit current increases when decreasing cavity thickness, as shown in Figure 4b for two different cavity dielectrics. This increase in current with decreasing cavity thickness corresponds to an increasing range of suppressed optical modes with decreasing thickness, as described in the Introduction.

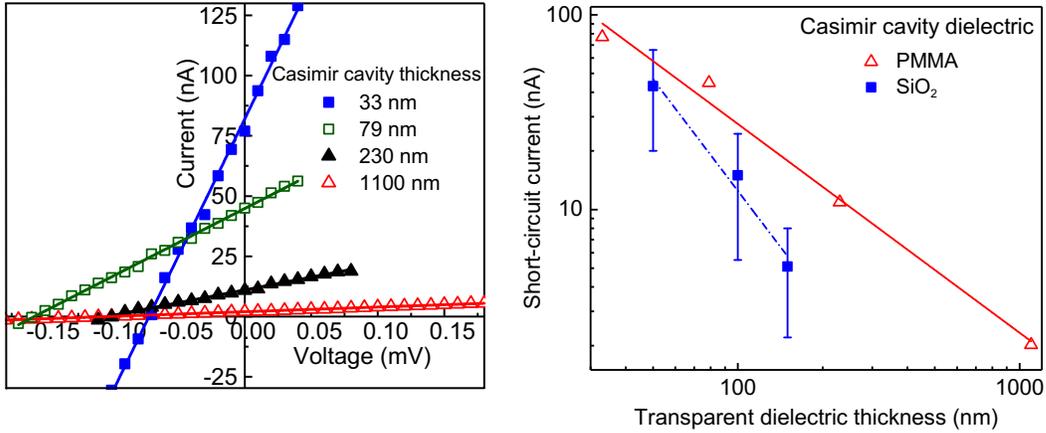

**Figure 4.** Electrical response as a function of cavity thickness. (**a**) Current as a function of voltage for different polymethyl methacrylate (PMMA) cavity thicknesses. (**b**) Short-circuit current as a function of cavity thickness for PMMA and SiO$_2$–filled cavities.

To understand the current-producing mechanism, we carried out tests to determine whether there is evidence that it involves hot charge carriers generated from optical fields in the optical cavity. These carriers, generated in the Pd upper electrode near the interface to the transparent dielectric cavity shown in Figure 1, could be injected into the insulator if they could traverse the Pd without being scattered. This photoinjection current should decrease when increasing the upper electrode thickness because of the increased scattering of the hot carriers before they reach the insulator, which results in excited carriers not contributing to the current. The hot electron mean-free path length in metals at room temperature is on the order of 10 nm [14]. On the other hand, when the upper electrode thicknesses is below the absorption depth of Pd, which is 10 nm for 0.4 μm radiation [15], the photoinjection current would be expected to decrease when decreasing the electrode thickness because the rate of carrier excitation is reduced. Figure 5a shows the short circuit current as a function of upper electrode thickness. This current does, in fact, decrease with increasing thickness, and also decreases for the thickness below 10 nm, peaking for a Pd thickness of approximately 12 nm.

To be collected, these hot carriers must traverse the insulator either ballistically or via tunneling. The ballistic transport is limited by the mean-free path length in the insulator, exponentially with insulator thickness [17]. In either case, the short-circuit current would be expected to decrease with increasing insulator thickness if the current is due to charge injection through the insulator. This trend is observed by the data of Figure 5b.

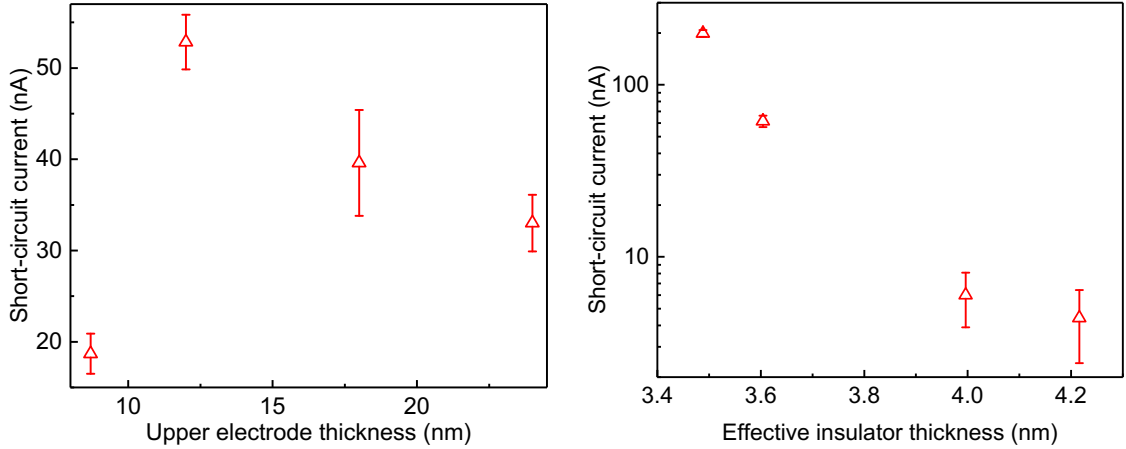

**Figure 5.** Tests for photoinjection of hot carriers through the Pd upper electrode and through the insulator. (**a**) Short-circuit current as a function of upper electrode Pd thickness. (**b**) Short-circuit current as a function of effective insulator thickness. Both trends are consistent with hot-carrier photoinjection from optical fields in the optical cavity.

Although the devices of Figure 4 were fabricated using the GSM process and the devices of Figure 5 used standard photolithography, devices fabricated by both processes exhibited similar trends. The GSM process allowed for much shorter fabrication times and for the absence of residual photoresist (described in Section 2.2), but it did not allow for large device areas or for varying the device area.

*3.2    Testing for Experimental Artifacts*
3.2.1    Stability over Time
We carried out a series of experiments to test whether the results presented above might be due to some sort of experimental artifact rather than a genuine electrical response from these structures. One concern is whether the observed current is a transient or hysteresis effect, possibly due to charging, as opposed to being a stable output from the devices. If, for example, one charge was trapped for every $Al_2O_3$ molecule (having a lattice constant of 0.5 nm) in a 2.5-nm thick layer of area 0.02 μm$^2$, depleting that charge could produce a current of 20 nA for 3.2 μs. To test for this, we measured the short-circuit current continuously over a period of four hours. The data, provided in Figure 6a, shows no change over time.

3.2.2    Area Dependence
Another concern is whether the output current is collected from just the active area covered by the upper electrode, or whether it is due to some other effect that would not scale with the active area. To test for this, we fabricated devices with a range of areas, in which the overlap shown in Figure 3 was varied. The results are shown in Figure 6b. The current scales linearly with the active area, supporting the conclusion that the source for the current is the active area.

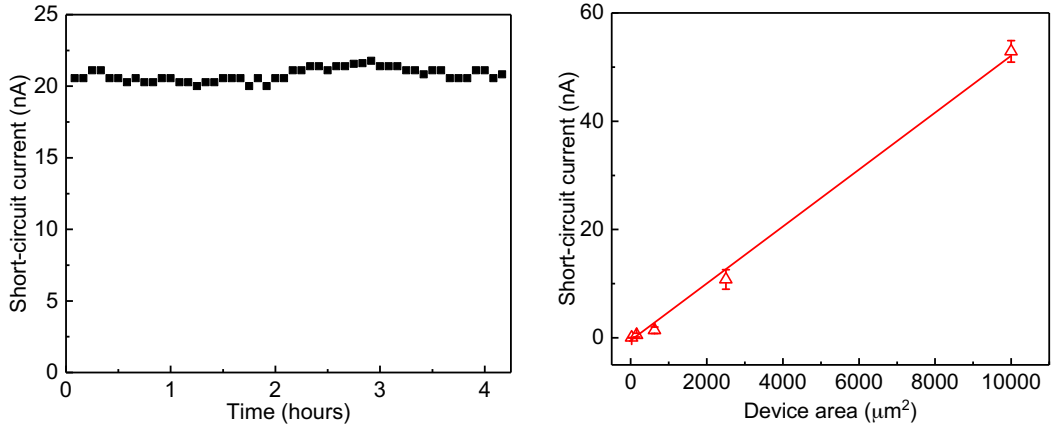

**Figure 6.** Two tests to check whether the measured current could be an experimental artifact. (**a**) Short- circuit current as a function of time over a period of four hours for a GSM device. (**b**) Short-circuit current as a function of active device area, as defined by the upper electrode area, for devices fabricated using the photolithography process.

3.2.3  Array Dependence

In addition to scaling with area, the short-circuit current should scale with the number of devices in parallel. Similarly, the open-circuit voltage should scale with the number of devices in series. If, for example, the output was the result of thermoelectric effects at the contacts, it would not scale with the number of devices. We tested for this possibility by fabricating and measuring two types of 4 × 4 arrays. One is a staggered array, schematically depicted in Figure 7a. It is a combination of series and parallel connections designed to reduce the effect of defective individual devices. The other 4 × 4 array is a series–parallel array consisting of four parallel sets of four devices in series. The results are shown in Figure 7b. The array currents and voltages are approximately four times those of the single devices, supporting the validity of the measured results.

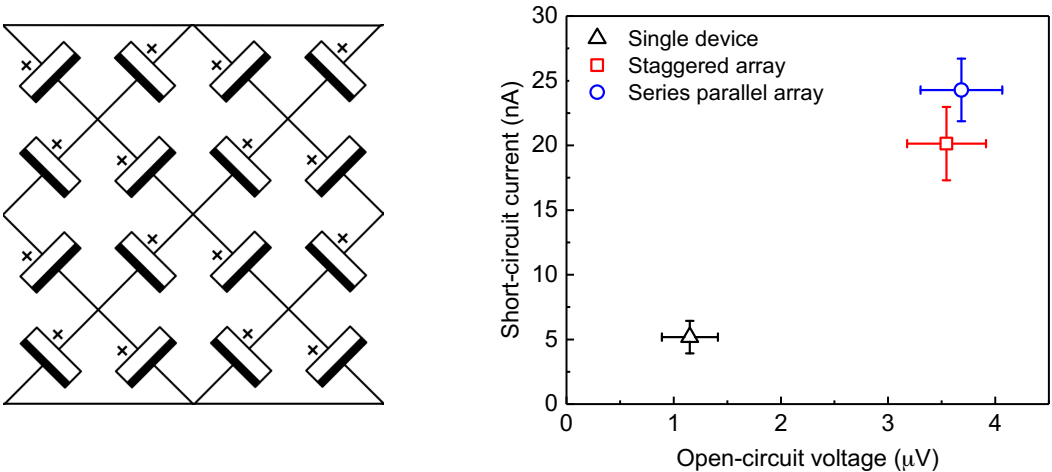

**Figure 7.** Measurements of device arrays. (**a**) Schematic representation of a 4 × 4 staggered array, where the circuit symbols represent

cavity/electronic-device elements; the measured total short- circuit current between the top and bottom bars and the open-circuit voltage are each four times that for a single element. (**b**) Measured short-circuit current and open-circuit voltage for single devices, staggered 4 × 4 arrays, and series–parallel arrays consisting of four parallel sets of four devices in series.

3.2.4   Processing Dependence

From experience fabricating and testing many thousands of MIM structures, we are confident that it is not the MIM structure alone that produces the observed electrical output. It is conceivable, however, that it is the additional processing of the MIM structures to form the adjoining cavity, and not the cavity itself, that gives rise to the output. To check for that, we measured devices at different stages of cavity formation. Figure 8 shows the short-circuit current from MIM devices at these different stages. The first stage is for an as-built MIM structure. Only a negligible current is produced. We then annealed the MIM structure at 180 ◦C for 15 min to replicate the temperature cycle that it would undergo during the process in which the mirror was defined. Again, no significant current was evident. We then deposited the cavity dielectric, i.e., PMMA in one case and $SiO_2$ in the other, and we saw no change in the current. Finally, after depositing the aluminum mirror, the current jumped to the values that we observed in the completed devices. This makes it clear that it is the cavity with the mirror, and not just the device processing, that yields a device producing the observed electrical output.

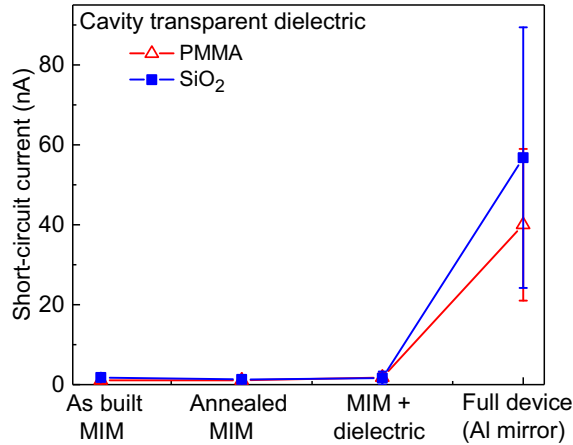

**Figure 8.** Effect of cavity formation on short-circuit current. The anneal was carried out at 180 °C for 15 min to replicate the mirror processing temperature cycle. Only a completed device produces a significant current.

The characteristics remain stable over time; for instance, for the full devices of Figure 8, after approximately six months the PMMA-cavity device output degraded by less than 10% and the $SiO_2$-cavity device output degraded by less than 20%.

3.2.5   Current Leakage through the Cavity

A possible source for the observed currents and voltages produced between the MIM electrodes might be leakage currents through the transparent dielectric from the mirror, which somehow picks up anomalous voltages. For this to be the case, resistance between the mirror and the MIM upper electrode would have to be on the order of or less than the resistance between the MIM electrodes. To test for this, we measured the relevant resistances in some completed devices. The results, given in Figure 9, show that the observed currents could not be due to leakage from the mirror.

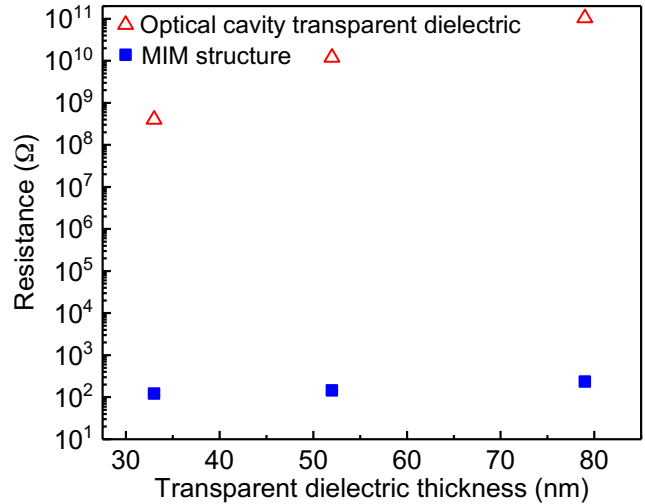

**Figure 9.** Comparison of resistance values across the optical cavity transparent dielectrics and across MIM structures. Because of the much higher resistance between the mirror and upper electrode than between the upper and the base electrodes (shown in Figure 1), the currents observed between the electrodes could not be the result of leakage from the mirror.

3.2.6    Electromagnetic Pickup

Another potential source of anomalous currents and voltage is ambient electromagnetic radiation. Electromagnetic pickup might occur somewhere in the device and result in current through the MIM structure, which would rectify it to produce the $I(V)$ characteristics shown in Figure 4a. To avoid such rectification, we designed the MIM devices to have low barrier heights and consequently linear $I(V)$ characteristics, as is evident from the lack of curvature in the data of Figure 4a. Still, a slight nonlinearity might rectify picked up signals. To test for that, we carried out measurements in three different environments: (i) the usual measurement conditions using a probed wafer chip mounted onto a measurement stage exposed to ambient fields, (ii) inside a mu-metal box, which blocks low frequency electromagnetic radiation, and (iii) inside an aluminum box, which blocks higher frequency radiation. The results, given in Figure 10, show that the current–voltage characteristics are the same for all three measurements. While these environments do not totally block all ambient radiation, the fact that the three measurements give the same results make it highly unlikely that electromagnetic pickup is the source for the electrical outputs that we observed.

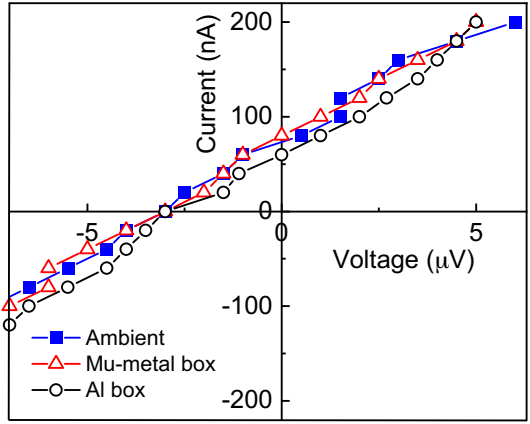

**Figure 10.** Effect of blocking ambient electromagnetic radiation during measurement of current-voltage characteristics. The characteristics measured under open ambient conditions did not change when the device was place in mu-metal or aluminum boxes.

3.2.7   Thermoelectric Effects on Electrodes

A common source of errors in low-voltage measurements is thermoelectric effects. One type is generated from a temperature difference between electrodes of different types at the device and at the measurement electronics. This voltage can be cancelled using the voltage reversal method [12] during measurement, as described in Section 2.1. We used this method consistently in our measurements. Another experiment that would indicate whether this type of thermoelectric effect could be the source of the electrical output is the measurement of device arrays. As the devices on a single substrate and at a uniform temperature are linked together, the thermoelectric voltage measured at the electronics would not change, and so the voltage would not scale with the number of devices in series. As shown in Figure 7b, the measured voltage does scale with the number of devices in series, and so it is not due to such a thermoelectric effect. Both the voltage reversal method and the device array results show that this type of thermoelectric voltage does not affect the results.

3.2.8   Thermoelectric Effects on Devices

Another potential source of thermoelectric effects is temperature differences within the sample itself. If the upper electrode was at a slightly different temperature than the lower electrode, this would generate a thermoelectric voltage. Such a temperature difference would be unlikely because of the much greater thermal conductance across the thin insulator than between the upper part of the device, shown in Figure 1, and the surrounding air. To be sure that this sort of thermoelectric effect is not the source of the measured electrical output, we carried out measurements to test for effects from such a temperature difference. The wafer chip containing the device is held tightly onto the metal measurement stage with a vacuum chuck. We varied the temperature of the measurement stage while the ambient temperature remained constant, and measured the short-circuit current and open-circuit voltage from the device. If there was a difference in temperatures between the upper and lower MIM electrodes that gave rise to a thermoelectric voltage, such a test would shift it or reverse its polarity. No change in the output voltage or current was observed, as shown in Figure 11, providing evidence that a temperature difference is not the source of the measured electrical output.

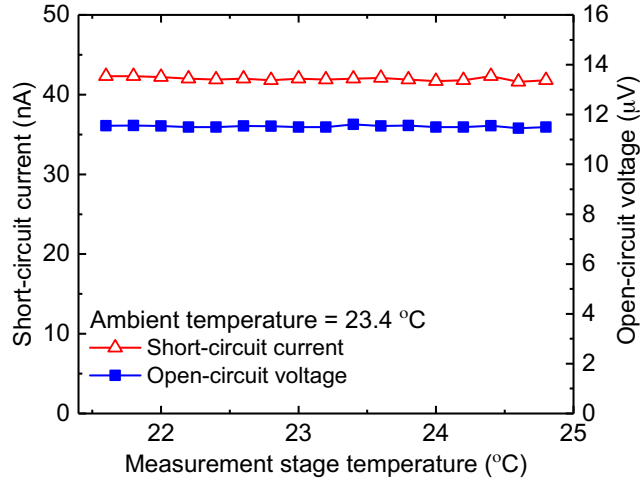

**Figure 11.** Test for possible thermoelectric effects. The electrical output is measured as the difference between the substrate and ambient temperatures is varied. The measured short-circuit current and open-circuit voltage does not vary, providing evidence that such a temperature difference is not the source of the measured electrical output.

The data shown here are not flukes observed in rare devices. The trends reported in this paper were replicated in over 1000 MIM-based devices produced in 21 different batches. Virtually all the devices with working (nonshorted) MIM structures exhibited the type of electrical characteristics shown in Figure 4a.

Based on all these checks for experimental artifacts, it appears that the measured electrical characteristics of the photoinjector cavity devices are, in fact, real and due to the devices themselves.

## 4. Discussion

The data show than when an MIM structure adjoins a thin optical cavity, electrical power is produced. An extensive range of tests show that real power is provided, and the results are not due to experimental artifacts. We consider a possible mechanism for the electrical characteristics based upon the experimental observations.

To produce current, the upper electrode in the MIM device must be thin, less than or on the order of a mean-free path length for ballistic charge carriers, consistent with Figure 5a. This suggests that the carriers are photoinjected from the cavity side of the MIM device. The observation that the current decreases with increasing insulator thickness, shown in Figure 5b, is consistent with the charge traversing the insulator by surmounting the metal/insulator barrier or tunneling through it.

The current increases with decreasing cavity thickness, as shown in Figure 4b. Optical cavities largely suppress wavelengths greater than twice the cavity spacing, such that the suppressed band extends to shorter wavelengths for thinner cavities. Therefore, the increase in current corresponds to an increasingly wide band of suppressed optical modes in the cavity. The source of these optical modes could be the quantum vacuum field, which gives rise to the Casimir force [18–21], the Lamb shift [22], and other physical effects [23]. It was argued that the use of energy from the vacuum field does not violate fundamental laws of thermodynamics [24].

In what follows, we present an operational model consistent with the observations to provide an interpretation of the results until a rigorous theoretical explanation is developed. The energy density of the quantum vacuum field varies with frequency cubed [23], and therefore the energy density of the suppressed cavity modes would vary with the reciprocal of the cavity thickness cubed. At first blush, one might expect to see this cubic dependence in Figure 4b. However, a multiplicity of other frequency-dependent processes could obscure this cubic dependence. They include (i) the dependence of photoinjection yield on photon energy, as described by extensions of Fowler's theory of photoemission [25]; (ii) limitations in the transport of high energy carriers through the Pd upper electrode due to the interband transition threshold [26]; (iii) the mirror energy-dependent reflectivity; (iv) the energy-dependent absorptivity of the transparent dielectric; and (v) the energy dependence of hot-carrier scattering [27]. A quantitative fit to the data of Figure 4b would require an extensive investigation of each of the energy-dependent mechanisms involved in producing the current. Despite this multiplicity of effects, the overall increase in current with decreasing cavity thickness is qualitatively consistent with what would be expected for the quantum vacuum field as the source for the hot-carrier excitation.

The hot carriers could be electrons, holes, or a combination of the two. The barrier heights are the main factor that determines which one dominates. The effective barrier heights for electrons between the Pd upper electrode and the $NiO_x$ and $Al_2O_3$ insulators are approximately 0.2 eV and 0.3 eV, respectively, whereas for holes, the respective barrier heights are 3.2 eV and 5.9 eV [6]. For the materials in the devices reported here, the higher barriers for holes are consistent with electrons being the dominant carriers. As described in the introduction, the charge transport through the insulator could be ballistic or via tunneling. For an insulator thickness below 4 nm, which is the case for the devices reported here, the dominant mechanism is tunneling [6].

The measured current from the devices is positive, i.e., in the direction of the arrow shown in the measurement circuit of Figure 1. This corresponds to a net current of electrons flowing from the nickel base electrode through the insulator to the palladium upper electrode. This direction of the current can be understood in terms of the three current components shown in Figure 12.

1. We consider first the MIM device in the absence of the optical cavity and the mirror. Component A is produced by free-space ambient optical modes impinging on the upper electrode, where they excite hot electrons. These hot electrons are injected into the insulator and then are absorbed in the base electrode. Component B is due to electrons excited within the upper electrode, e.g., from plasmonic zero-point fluctuations [28]. The electrons are injected into the insulator and then absorbed in the base electrode. Component C, in the opposite direction, is due to electrons that are excited by fluctuations within the base electrode, injected into the insulator, and then absorbed in the upper electrode. There is no optically excited current component of electrons from the base electrode to the upper electrode because the base electrode is thicker than the electron mean-free path length, and so any electrons excited at the outer (lower) surface of the electrode are scattered before reaching the insulator. In equilibrium, the net current is zero, and component C is balanced by the sum of components A and B.
2. We now consider the MIM device in the presence of an adjoining optical cavity and the mirror. The addition of the adjoining structure upsets the balance in current components. Because the cavity reduces the density of the optical modes impinging on the upper electrode, component A is reduced while components B and C remain unchanged. This results in a net electron current from the base electrode to the upper electrode, which is consistent with our observations.

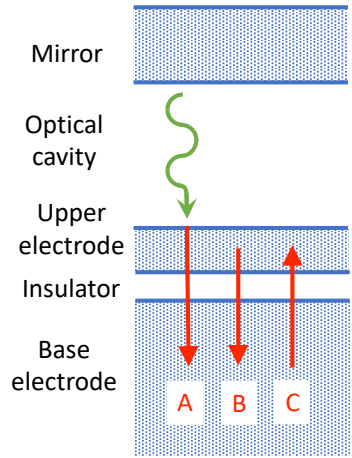

**Figure 12.** Cross section of the photoinjector device, showing optically generated electron current component **A**, and internally generated components **B** and **C**.

To understand the enhancement of the measured current with varying thicknesses shown in Figure 5a,b we again consider the current components of Figure 12. As discussed with respect to Figure 5a, decreasing the upper electrode thickness from 24 nm down to the absorption depth of ~10 nm, allows an increasing fraction of the photoexcited electrons to traverse the upper electrode without being scattered, with the result that they can produce measurable current. As discussed with respect to Figure 5b, decreasing the insulator thickness increases the fraction of photoexcited electrons that can traverse the insulator and produce measurable current. In both cases, decreasing the thickness of the layers tends to increase the proportion of incoming photons that excite electrons, which can then be injected. The effect is to enhance the photoinjection yield, which increases electron current component A. To understand how the increased photoinjection yield can result in a greater electron current in the direction opposite to the photoinjection when an adjoining cavity is added, we consider the process first without and then with the cavity.
1. In examining the balance of current components when the MIM device is not perturbed by the presence of the optical cavity and mirror, we consider first the device in the absence of the cavity structure. To maintain the zero net current, this increase in photoinjection yield with changing layer thickness must be balanced by a decrease in the internally generated component B or an increase in the internally generated component C. These changes in components B and C result from the thickness changes and are independent of whether there is an adjoining optical cavity or not.

2. We once again consider the MIM device in the presence of an adjoining optical cavity and mirror. With the reintroduction of this adjoining structure, the enhanced photoinjection yield represented by component A now leads to a greater suppression of component A. Because of the greater reduction in component A, the net electron current from the base electrode to the upper electrode is enhanced. Thus, increasing the photoinjection yield leads to an enhanced current that is induced by the presence of an adjoining optical cavity, in the direction opposite to the photoinjection current.

There are additional current components in play beyond those indicated by the three arrows in Figure 12, but they are not expected to add significantly to the current.
- Additional components result from the fact that the insulator itself forms a very thin optical cavity. This cavity is symmetric with respect to the MIM structure itself, as opposed to the optical cavity shown in the figure, which is to one side of the MIM structure. Because there is no longer a cavity having a reduced density of vacuum modes on one side of the tunneling region, the current components in each direction balance each other out, resulting in no net current. This is consistent with the observation that MIM structures without the adjoining optical cavity do not produce a current, as shown in Figure 8.
- Another component results from the upper electrode being slightly transparent. As a result, a small fraction of the optical radiation from the optical cavity impinges on the lower electrode and produces hot carriers. Because the optical transmission through the upper electrode is small, this produces only negligible effects.
- Additional effects, such as those from the surface plasmon modes in the cavity [29], cannot be ruled out.

There is a particular characteristic of MIM structures that adjoin the optical cavities that may be key in allowing the observed current and voltage outputs to be induced. This characteristic is the time required for hot-charge carriers to traverse the combination of the upper electrode and the thin insulator, followed by capture in the base electrode. The entire process can be very fast. The hot-carrier velocity in the metal is at least the Fermi velocity of $10^6$ m/s [30,31]. For a metal thickness of approximately 10 nm, the resulting transit time is less than 10 fs. In the insulator, which is even thinner, if the carriers travel ballistically, the velocity is $10^6$ m/s [32]. This results in a roughly 1 fs transit time through the insulator. This is on the order of the same time that is required for tunneling [33], which appears to be the dominant transport mechanism, as discussed above. Finally, the hot carriers scatter inelastically in the base electrode, with a lifetime of at most 10 fs. The combination of hot-carrier transport and scattering takes place in the order of 10 fs.

We speculate that the reason the femtosecond transit and capture gives rise to the observed electrical output has to do with an uncertainty-principle-like relation that governs the process. It has been argued that an amount of energy $\Delta E$ may be borrowed from the quantum vacuum field for a time $\Delta t$ [34–36], although that has yet to be supported by experiments [37]. For $\Delta E \Delta t \sim \hbar/2$, hot electrons from 1 eV excitations would be available for 0.3 fs. For a transit and capture that is longer than that, a fraction of the hot electrons would be available. MIM devices adjoining optical cavities could capture photoinjected electrons sufficiently quickly and irreversibly, which would give rise to the observed currents.

## 5. Conclusions

In metal–insulator–metal (MIM) tunnel devices adjoining thin optical cavities, we consistently observed a small output current and voltage, as reproduced in over 1000 devices produced in 21 batches. When the cavities were made thinner, which corresponds to suppressing a wider range of optical modes, the current increased. The output scaled with number of devices in parallel and series, and the current scaled with the device area. Changing the layer thicknesses in the MIM structure resulted in changes in the current that are consistent with modifying the suppression of hot electron photoinjection from the side of the MIM structure adjoining the optical cavity.

We carried out a set of tests to determine whether the measured electrical output could be the result of some sort of experimental artifact. The results support the conclusion that the source of the electrical output is not due to measurement offsets or errors, transient stored charge, characteristics of the structure not related to the optical cavity, electromagnetic pick-up, electrical leakage through the optical cavity, or thermoelectric effects in the electrodes or in the device itself. All evidence is that the device itself produces the measured outputs.

The observations are consistent with the optical cavity upsetting an equilibrium balance of currents in the MIM structures. We posit that quantum fluctuations excite the observed currents. If access to such excitation is limited by a $\Delta E \Delta t$ uncertainty-principle-like relation, the available energy $\Delta E$ would be accessible for a very short time $\Delta t$. The ultra-fast charge transport and capture in MIM devices is compatible with such a short time requirement.

**Author Contributions:** G.M. conceived the device concept, directed the research, and wrote the manuscript; A.W. directed the fabrication and testing, and carried out much of the GSM device fabrication; D.D. carried out fabrication of the initial devices and layer depositions; and D.B. carried out design and fabrication of the photolithographic devices. All authors have read and agreed to the published version of the manuscript.

**Funding:** Much of the work was supported by The Denver Foundation, which we gratefully acknowledge.

**Institutional Review:** Not applicable.

**Informed Consent Statement:** Not applicable.

**Data Availability Statement:** All data analyzed for this study are included in this published article.

**Acknowledgements:** We greatly appreciate the insightful comments on our results from P. C. W. Davies, E. A. Cornell, J. Maclay, and M. J. Estes, and M. McConnell and R. Cantwell for help in verifying and improving the measurements, and additional comments on the manuscript from B. Pelz and J. Stearns.